\documentclass[aps,prl,showpacs,twocolumn,preprintnumbers,superscriptaddress,nofootinbib,floatfix,10pt]{revtex4-1}
\usepackage{amsfonts,amssymb,stmaryrd,latexsym,amsmath,braket}
\usepackage{graphicx,subfigure}
\usepackage{comment}
\usepackage{times}
\usepackage{slashed}
\usepackage{bm}
\usepackage{braket}
\usepackage{chapterbib}
\usepackage{color}

\newcommand{\beginsupplement}{%
        \setcounter{table}{0}
        \renewcommand{\thetable}{S\arabic{table}}%
        \setcounter{figure}{0}
        \renewcommand{\thefigure}{S\arabic{figure}}%
        \setcounter{equation}{0}
        \renewcommand{\theequation}{S\arabic{equation}}%
     }

\makeatletter
\let\saved@includegraphics\includegraphics
\AtBeginDocument{\let\includegraphics\saved@includegraphics}
\makeatother

\begin{document}

\title{\textit{Ab initio} study of nuclear clustering in hot dilute nuclear matter}


\author{Zhengxue Ren}
\email{z.ren@fz-juelich.de (corresponding author)}
\affiliation{Institut f\"{u}r Kernphysik, Institute for Advanced Simulation and J\"{u}lich Center for Hadron Physics, Forschungszentrum J\"{u}lich, D-52425 J\"{u}lich, Germany}
\affiliation{Helmholtz-Institut f\"{u}r Strahlen- und Kernphysik and Bethe Center for Theoretical Physics, Universit\"{a}t Bonn, D-53115 Bonn, Germany}

\author{Serdar Elhatisari}
\email{selhatisari@gmail.com}
\affiliation{Faculty of Natural Sciences and Engineering, Gaziantep Islam Science and Technology University, Gaziantep 27010, Turkey}
\affiliation{Helmholtz-Institut f\"{u}r Strahlen- und Kernphysik and Bethe Center for Theoretical Physics, Universit\"{a}t Bonn, D-53115 Bonn, Germany}

\author{Timo A. L\"{a}hde}
\email{t.lahde@fz-juelich.de}
\affiliation{Institut f\"{u}r Kernphysik, Institute for Advanced Simulation and J\"{u}lich Center for Hadron Physics, Forschungszentrum J\"{u}lich, D-52425 J\"{u}lich, Germany}
\affiliation{Center for Advanced Simulation and Analytics (CASA),~Forschungszentrum~J\"{u}lich,~D-52425~J\"{u}lich,~Germany}

\author{Dean Lee}
\email{leed@frib.msu.edu}
\affiliation{Facility for Rare Isotope Beams and Department of Physics and Astronomy, Michigan State University, East Lansing, MI 48824, USA}

\author{Ulf-G.~Mei{\ss}ner}
\email{meissner@hiskp.uni-bonn.de}
\affiliation{Helmholtz-Institut f\"{u}r Strahlen- und Kernphysik and Bethe Center for Theoretical Physics, Universit\"{a}t Bonn, D-53115 Bonn, Germany}
\affiliation{Institut f\"{u}r Kernphysik, Institute for Advanced Simulation and J\"{u}lich Center for Hadron Physics, Forschungszentrum J\"{u}lich, D-52425 J\"{u}lich, Germany}
\affiliation{Tbilisi State University, 0186 Tbilisi, Georgia}

\begin{abstract}
We present a systematic \textit{ab initio} study of clustering in hot dilute nuclear matter using nuclear lattice effective
field theory with an SU(4)-symmetric interaction.   We introduce a method called light-cluster distillation to determine the abundances of dimers, trimers, and alpha clusters as a function of density and temperature.  Our lattice results are compared with an ideal gas model composed of free nucleons and clusters. Excellent agreement is found at very low density, while deviations from ideal gas abundances appear at increasing density due to cluster-nucleon and cluster-cluster interactions. In addition to determining the composition of hot dilute nuclear matter as a function of density and temperature, the lattice calculations also serve as benchmarks for virial expansion calculations, statistical models, and transport models of fragmentation and clustering in nucleus-nucleus collisions.
\end{abstract}
\maketitle
\date{today}

The equation of state of nuclear matter and its composition are topics that lie at the heart of nuclear physics and determine important properties of neutron star evolution, binary neutron star mergers, core-collapse supernovae, and other astrophysical processes.  Many terrestrial studies have used particle yields and cluster distributions from heavy-ion collisions to infer the equation of state and composition of nuclear matter as a function of density, temperature, and isospin \cite{Iglio:2005ve,Henzlova:2005ed,Souliotis:2006dm,Tsang:2006zc,Zhang:2007hmv,Mocko:2008rj,Huang:2010jp,Hagel:2011ws,Qin:2011qp,Ono:2019jxm,Pais:2019jst,Sorensen:2023zkk}.  In this work, we focus on the properties of hot dilute nuclear matter at densities below saturation and temperatures below the pion production threshold.  This regime is relevant to the liquid-gas phase transition in nuclear matter~\cite{Typel:2009sy} and core-collapse supernovae~\cite{Arcones:2008kv, Sumiyoshi:2008qv}.  While previous {\it ab initio} lattice calculations have determined the equation of state and phase diagram \cite{Lu:2019nbg}, in this work we present the first {\it ab initio} calculations of the cluster composition of hot dilute nuclear matter.

Nuclear clustering in hot dilute nuclear matter has been studied using numerous methods, including virial expansions~\cite{Horowitz:2005nd, OConnor:2007kup}, relativistic mean-field theory (combined with other approaches)~\cite{Shen:1998gq, Typel:2009sy, Qin:2011qp, Pais:2019jst}, statistical models~\cite{Henzlova:2005ed, Iglio:2005ve, Souliotis:2006dm, Tsang:2006zc, Ropke:2008qk, Typel:2009sy, Qin:2011qp, Hagel:2011ws}, and transport models~\cite{Zhang:2007hmv, Mocko:2008rj, Huang:2010jp, Ono:2019jxm}.  The phenomenon of nuclear clustering is also important for understanding the excitation spectra of light nuclei ~\cite{Freer:2017gip} and, possibly, also heavy nuclei~\cite{Souza:2023kor}.  Recently, nuclear clustering in dilute nuclear matter has been proposed as an explanation for alpha formation in alpha-decay~\cite{Tanaka:2021oll} and ternary fission~\cite{Ren:2021hoy}.  In spite of recent progress using various \emph{ab initio} (first principles or fully microscopic) methods~\cite{Stroberg:2016ung, Piarulli:2017dwd, Lonardoni:2017hgs, Gysbers:2019uyb, Smirnova:2019yiq, Contessi:2017rww}, the treatment of non-zero temperature is difficult.  Most studies have therefore relied on many-body perturbation theory \cite{Baldo:1999cvh, Holt:2013fwa, Soma:2009pf, Carbone:2018kji, Carbone:2019pkr, Carbone:2019hym, Yuksel:2022ezb}.  However, the formation of clusters is a signature of strong many-body correlations and is therefore difficult to produce entirely from perturbation theory starting from mean field theory.

Nuclear lattice effective field theory (NLEFT)~\cite{Lee:2008fa, Lahde2019book} is a powerful numerical method which uses finite volume Monte Carlo (MC) methods and is formulated in the framework  of nuclear chiral EFT~\cite{Epelbaum:2008ga}.  As the NLEFT simulations include many-body correlations to all orders, effects such as deformation and clustering appear naturally.  NLEFT defines a systematically improvable \textit{ab initio} nuclear theory starting from bare nuclear forces and has favorable computational scaling with the number of nucleons. NLEFT has been successfully applied to probe alpha clustering in finite nuclei~\cite{Epelbaum:2011md, Epelbaum:2012qn, Epelbaum:2012iu, Epelbaum:2013paa, Elhatisari:2016owd, Elhatisari:2017eno, Shen:2021kqr, Shen:2022bak}.  In Ref.~\cite{Lu:2019nbg}, it was extended to calculations of the thermodynamics of nuclear systems using the pinhole trace algorithm (PTA).

In the present work, we use NLEFT simulations to determine the fractions of nucleons forming two-body, three-body, and four-body (alpha) clusters in symmetric nuclear matter as a function of temperature $T$ and density $\rho$.
We use the Wigner SU(4)-symmetric nuclear force introduced in Ref.~\cite{Lu:2018bat}, where the interaction is independent of spin and isospin. While the extension to high-fidelity chiral nuclear forces is straightforward, the calculations require new algorithms to perform perturbation theory for wave functions currently being developed \cite{Ma2023ro} and will therefore be presented in future work. It should also be noted that SU(4)-symmetric forces already provide a highly successful description for the ground-state properties of many light and medium-mass nuclei~\cite{Lu:2018bat}, for pure neutron matter and symmetric nuclear matter~\cite{Lu:2019nbg} as well as for the low-lying excited states of the $^{12}{\rm C}$ nucleus~\cite{Shen:2021kqr, Shen:2022bak}, where alpha clustering effects are extremely important.
Details about the Wigner SU(4)-symmetric nuclear interactions and the PTA can be found in the Supplemental Material~\cite{SM} and Refs.~\cite{Lu:2018bat, Lu:2019nbg}.  We note that the SU(4)-invariant deuteron is degenerate with the di-neutron and di-proton ground state and has less than half of the physical deuteron binding energy. The SU(4) symmetry makes the enumeration of light clusters simple and therefore useful for introducing the light-cluster distillation method.

We regard the clustering of nucleons in nuclear matter as a spatial localization of nucleons.  As a detailed probe of the nucleon correlations, we compute the following set of
correlation functions:
\begin{subequations}\label{eq:correlators}
  \begin{align}
   G_{11}(n) &=L^3\sum_{\{\sigma_i\tau_i\}}:\rho_{\sigma_1\tau_1}(0)\rho_{\sigma_2\tau_2}(\bm{n}):,\\
   G_{21}(n) &=L^3\sum_{\{\sigma_i\tau_i\}}:\rho_{\sigma_1\tau_1}(0)\rho_{\sigma_2\tau_2}(0)\rho_{\sigma_3\tau_3}(\bm{n}):,\\
   G_{31}(n) &=L^3\sum_{\{\sigma_i\tau_i\}}:\rho_{\sigma_1\tau_1}(0)\rho_{\sigma_2\tau_2}(0)\rho_{\sigma_3\tau_3}(0)\rho_{\sigma_4\tau_4}(\bm{n}):,\\
   G_{22}(n) &=L^3\sum_{\{\sigma_i\tau_i\}}:\rho_{\sigma_1\tau_1}(0)\rho_{\sigma_2\tau_2}(0)\rho_{\sigma_3\tau_3}(\bm{n})\rho_{\sigma_4\tau_4}(\bm{n}):.
 \end{align}
\end{subequations}
Here, $\bm{n}$ is a site on a cubic spatial lattice with lattice spacing $a$ and volume $L^3$, $n$ is the radial distance from the origin, the columns :: denote normal ordering, and
$\rho_{\sigma_i\tau_i}(\bm{n}) = a^+_{\sigma_i\tau_i}(\bm{n}) a_{\sigma_i\tau_i}(\bm{n})$ is the nucleon density operator for the spin $\sigma_i$ and isospin $\tau_i$.
By $\sum_{\{\sigma_i\tau_i\}}$, we denote summation over all spin-isospin indices without repetition of indices.
This avoids the strong short-range exclusion effects due to the Pauli principle.
The correlation functions in Eq.~\eqref{eq:correlators} are computed using the rank-one operator
method introduced in Ref.~\cite{Ma2023ro}.

\begin{figure}[!htbp]
  \centering
  \includegraphics[width=0.50\textwidth]{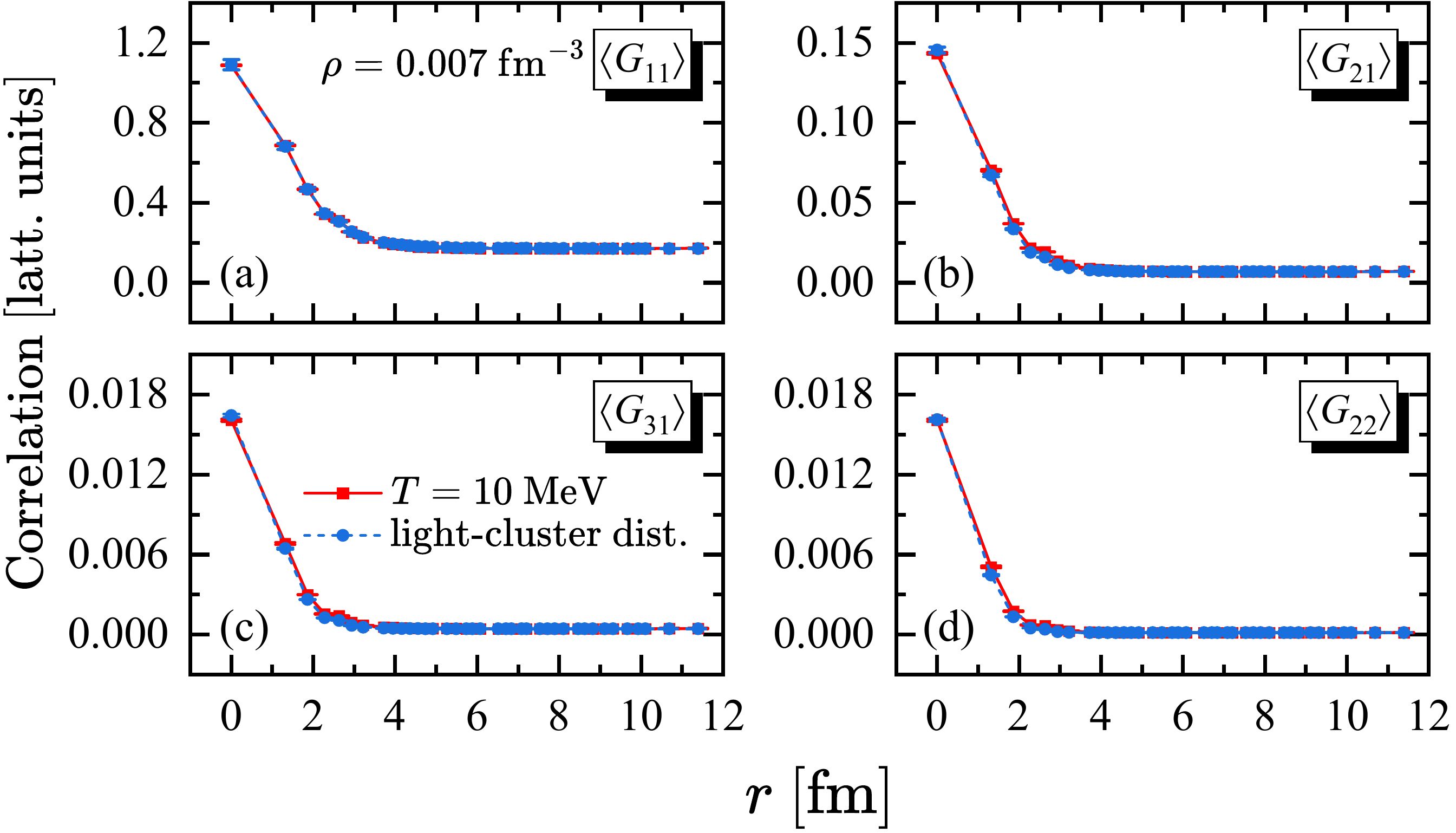}\\
  \caption{Correlation functions in Eq.~(\ref{eq:correlators}) versus relative distance $r$ in units of ${\rm fm}$. The solid lines and red squares show our results for the density $\rho = 0.007~{\rm fm}^{-3}$ (total particle number $A=16$ for the box size $L = 10$) and temperature $T=10~{\rm MeV}$.
  The dashed lines and blue circles show the fitted results of light-cluster distillation using Eq.~\eqref{eq:fit1}.
  The lines connecting the data points are intended to guide the eye.
  \label{fig1}}
  \end{figure}

To illustrate the characteristic features of the correlation functions given in Eq.~\eqref{eq:correlators} we consider a system with $A=16$ nucleons in a cubic box of length $L = 10$ corresponding to a density $\rho=0.007~{\rm fm}^{-3}$ and perform lattice simulations for temperature $T=10~{\rm MeV}$.
In Fig.~\ref{fig1} the lattice results are shown by the solid lines and red-square points.
We find that all the correlation functions are nearly flat for large spatial separations and have a strong peak at shorter distances.  As we will now see, this signal is consistent with a gas of small clusters.

In order to quantitatively determine the cluster abundances, we use the method of light-cluster distillation.  We write the correlation functions for hot dilute nuclear matter as a weighted sum of correlation functions for individual light clusters,
\begin{subequations}
\label{eq:fit1}
   \begin{align}
      \langle G_{11}(n)\rangle \approx &~ \langle G_{11}^{ l}\rangle +w_4\langle G_{11}(n)\rangle_4+ w_3\langle G_{11}(n)\rangle_3\nonumber \\
      &+w_2\langle G_{11}(n)\rangle_2,\\
      \langle G_{21}(n)\rangle \approx &~\langle G_{21}^{l}\rangle+w_4\langle G_{21}(n)\rangle_4 + w_3\langle G_{21}(n)\rangle_3,\\
      \langle G_{31}(n)\rangle \approx &~\langle G_{31}^{l}\rangle+w_4\langle G_{31}(n)\rangle_4,\\
      \langle G_{22}(n)\rangle \approx &~\langle G_{22}^{l} \rangle +w_4\langle G_{22}(n)\rangle_4.
   \end{align}
\end{subequations}
Here we narrow our focus to light clusters such as dimers ($^2{\rm H}$, $pp$, $nn$), trimers ($^3{\rm H}$, $^3{\rm He}$) and alpha particles ($^4{\rm He}$).  As we consider an SU(4)-invariant Hamiltonian and symmetric nuclear matter, we need not distinguish different types of dimers and trimers.
In Eq.~(\ref{eq:fit1}), $\langle\cdots\rangle_k$ denotes the computed correlation functions for dimers ($k=2$), trimers ($k=3$), and alphas ($k=4$) in their ground states.  The semi-positive valued parameters $w_k$ are interpreted as the number of clusters and determined by a least-square fitting procedure.  Finally, $G_{ij}^l$ denotes the long-range parts of the correlation functions defined by, e.g.,
\begin{equation}
  G_{11}^{l} = \frac{L^3}{\sum_{|\bm{n}'|>n_m}}\sum_{\{\sigma_i\tau_i\}}\sum_{|\bm{n}'|>n_m}:\rho_{\sigma_1\tau_1}(0)\rho_{\sigma_2\tau_2}(\bm{n}'):,
\end{equation}
where $n_m=8~{\rm fm}/a$ separates the short-range and long-range physics. For sufficiently large values of $n_m$, the results are independent of $n_m$.  Although the character of the density correlations at short distances depend on the regulator used for the nuclear interactions,  the distillation coefficients into clusters should be independent of the regulator when the low-energy nuclear interactions remain the same.

In Fig.~\ref{fig1}, the results from fitting lattice data to Eq.~(\ref{eq:fit1}), shown by the dashed lines and blue-circle points, indicate that the correlation functions are accurately represented by light-cluster distillation with the three parameters, $w_2$, $w_3$ and $w_4$. This finding suggests that the system is a gas of nucleons and light clusters.  After determining the $w_k$ parameters from light-cluster distillation, we compute the mass fractions given by $X_k = kw_k/A$ for clusters composed of $k$ nucleons as well as the total fraction of nucleons bound in light clusters defined as $X_{\rm bound} = X_2+X_3+X_4$.

We perform lattice simulations at temperatures ranging from $10$ to  $20~{\rm MeV}$ for densities $\rho = 0.007$ and $0.014~{\rm fm}^{-3}$, and we employ light-cluster distillation.  The heavier-cluster mass fraction with $A > 4$ is estimated to reach about $10\%$ at density $0.01~{\rm fm}^{-3}$ and temperature $10~{\rm MeV}$~\cite{Horowitz:2005nd}.  The abundance of heavier clusters with $A > 4$ are much smaller for lower densities and/or higher temperatures.  Correlation functions for $\rho = 0.007$ and $0.014~{\rm fm}^{-3}$ at several representative temperatures are shown in the Supplemental Material~\cite{SM}.

\begin{figure}[!htbp]
  \centering
  \includegraphics[width=0.48\textwidth]{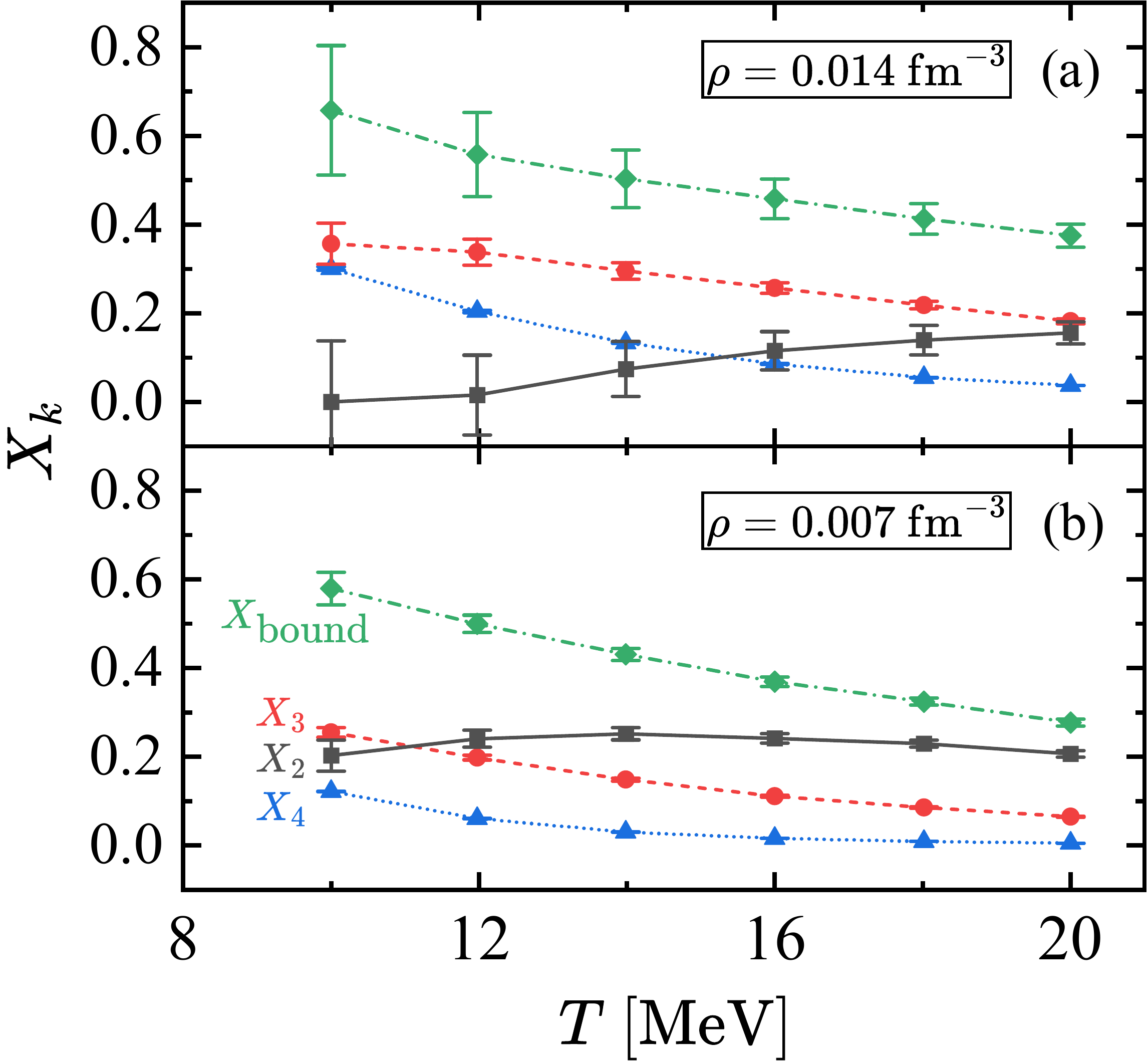}
  \caption{Mass fractions $X_k(\rho, T)$ for clusters with $k$ nucleons. Top panel:
  $\rho = 0.014~{\rm fm}^{-3}$.  Bottom panel: $\rho = 0.007~{\rm fm}^{-3}$. All results are for $L = 10$
  (in lattice units).
  The error bars reflect the fitting uncertainty for the $w_k$ in light-cluster distillation.
      \label{fig2}}
\end{figure}

In Fig.~\ref{fig2}, we show the mass fraction and total fraction results for $\rho = 0.007~{\rm fm}^{-3}$ and $0.014~{\rm fm}^{-3}$.
We find that the fraction of nucleons bound in light clusters decreases monotonously with increasing $T$.
This observation strongly suggests the thermal evaporation of nucleons from the clusters.
On the other hand, the mass fractions of dimers, trimers and alphas show somewhat complementary behaviors.
For fixed $\rho$, $X_3$ and $X_4$ decrease with temperature from $10$~MeV to $20$~MeV, while the behavior for $X_2$ is slightly increasing or relatively flat. For fixed $T$, $X_3$ and $X_4$ increase with density from $0.007~{\rm fm}^{-3}$ to $0.014~{\rm fm}^{-3}$, while $X_2$ is decreasing.  We have also perform lattice simulations at the same temperatures and densities using a larger volume of length $L=12$.  We find that the mass fraction results are nearly the same, and the results are shown in the Supplemental Material~\cite{SM}. This indicates that the finite size effects are small and the lattice results are close to the thermodynamic limit.  This is not unexpected since the box length $L$ is significantly larger than the thermal wavelength $\lambda = \sqrt{2\pi/(mT)}$, which is a measure of the correlation length at non-zero temperature.

As seen in Fig.~\ref{fig2} at lower temperatures and higher densities, the uncertainties on the determination of the light-cluster mass fractions become larger.  This is caused by the increasing proportion of clusters composed of more than four nucleons.  The determination of cluster abundances at lower temperatures and higher densities requires enumerating clusters with $A > 4$.  This can be performed by calculating and analyzing correlation functions beyond those in Eq.~(\ref{eq:correlators}) and will be addressed in future studies.  See the Supplemental Material for further discussion of heavier clusters ~\cite{SM}.

\begin{figure}[!htbp]
  \centering
  \includegraphics[width=0.48\textwidth]{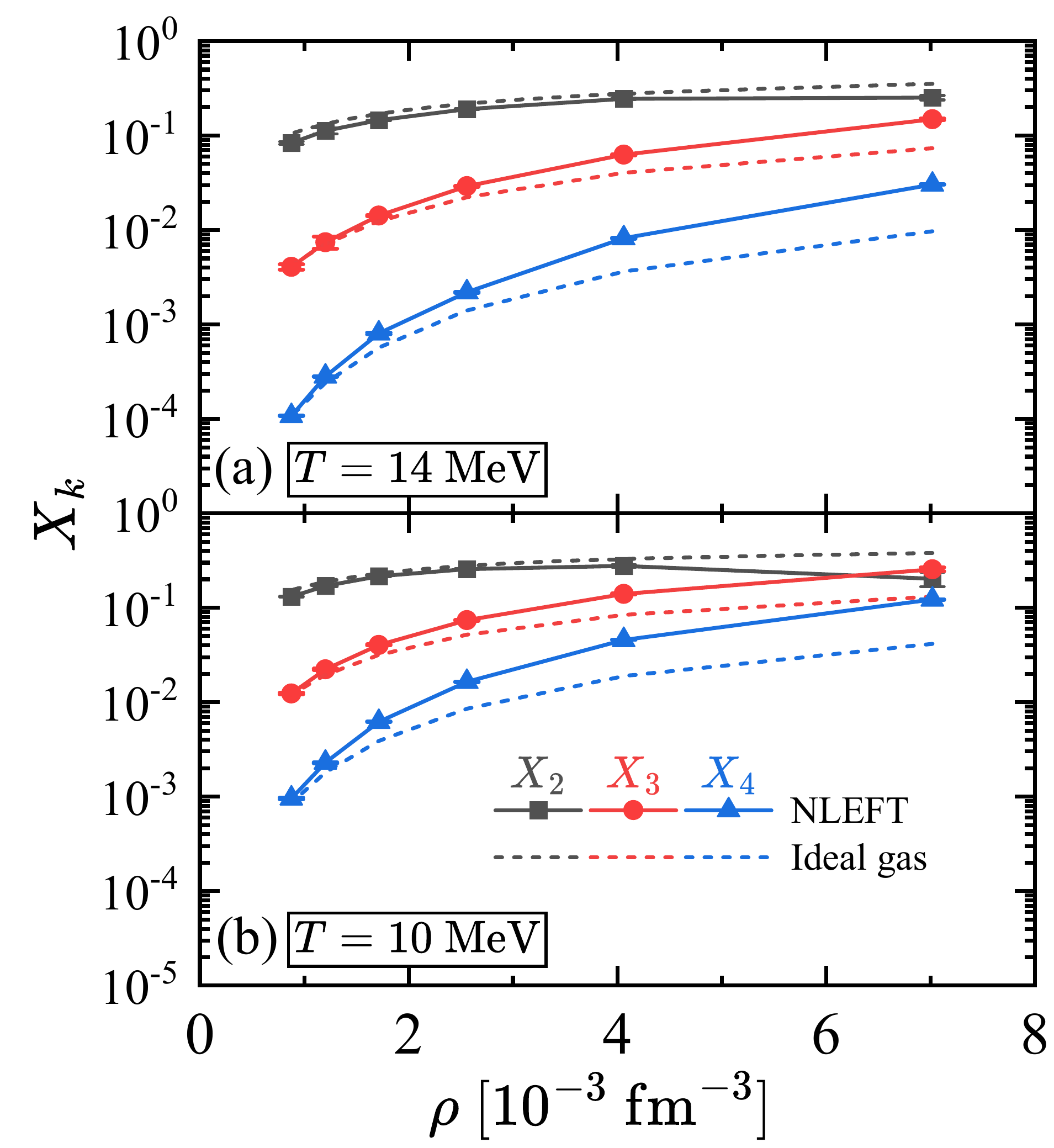}
  \caption{The mass fraction $X_k$ for temperatures $T = 14$ (top) and $10~{\rm MeV}$ (bottom) as a function of density.
  Results are shown for $A = 16$, and $L = 10 \ldots 20$ for steps of~$2$. The lattice results are shown by squares, circles, and triangles, while the dashed lines are the predictions from an ideal gas model.
  The error bars reflect the fitting uncertainty for the $w_k$ in light-cluster distillation.
  \label{fig3}}
  \end{figure}

In Fig.~\ref{fig3}, we compare our results for $X_k(\rho, T)$ with an ideal gas model, where the
system is assumed to consist of free nucleons, dimers, trimers, and tetramers (alpha clusters), governed by the
grand canonical ensemble~\cite{Pathria2011book, SM}. To ensure clarity and avoid any potential ambiguities arising from the presence of heavier clusters, we focus on densities $\rho \le 0.007~{\rm fm}^{-3}$ in our analysis.
Notable deviations become apparent around $\rho \simeq 0.007~{\rm fm}^{-3}$ for all types of clusters. Nevertheless, at lower densities, specifically around $\rho \simeq 0.001~{\rm fm}^{-3}$, the agreement is excellent.  These observations suggest that the deviations at higher $\rho$ arise from non-negligible cluster-nucleon and cluster-cluster interactions.

\begin{figure}[!htbp]
  \centering
  \includegraphics[width=0.48\textwidth]{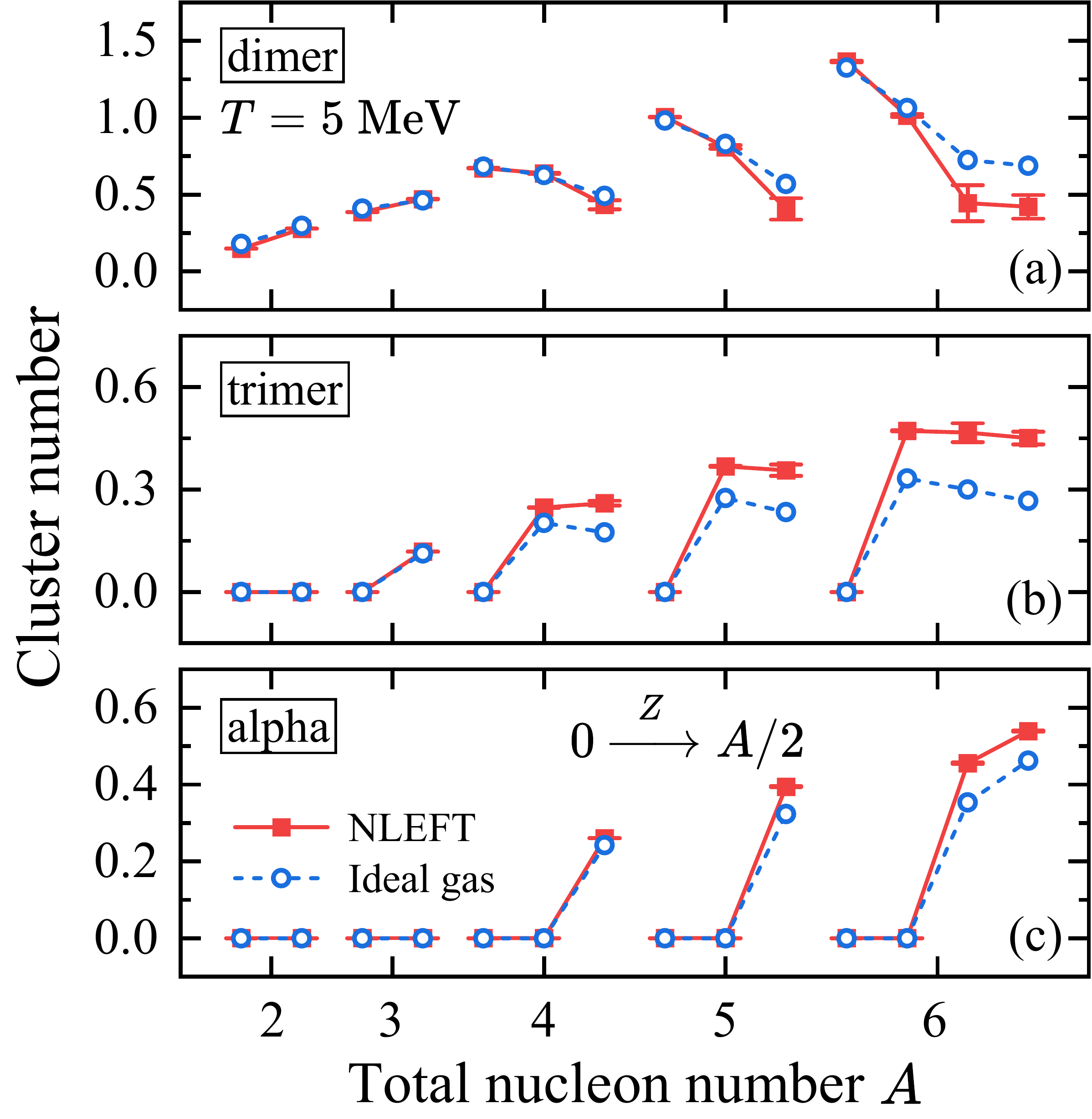}
  \caption{The number of dimers (top panel), trimers (middle panel), and alphas (bottom panel) as a function of $A$ and proton number $Z$.
  The lattice simulations are performed at $T=5~{\rm MeV}$ and for $L=10$.
  Circles with dashed lines show the predictions of an ideal gas model in the canonical ensemble.
  For each $A$, the proton number $Z$ is varied from $0$ (left) to $A/2$ (right).
  \label{fig4}}
\end{figure}

In order to quantitatively clarify such effects, we perform lattice simulations at temperature $T = 5$~MeV and for various numbers of nucleons $A$ in a cubic box of length $L=10$. For each $A$, we also vary the proton number $Z$ from $0$ to $A/2$ to step-by-step activate cluster-nucleon and cluster-cluster  interactions.  The results are presented in Fig.~\ref{fig4}.
The ideal gas results are predictions from the canonical ensemble with the same particle content and physical volume (for more details,
see~\cite{SM}).

We first consider the dimer case.
For $A \le 4$ and $Z \le 1$, the deviations between lattice and ideal gas results are negligible, which implies that nucleon-nucleon, nucleon-dimer, and dimer-dimer ($nn$--$nn$ and $nn$--$np$) interactions have no significant impact on the dimer abundance.
For $A = 4$ and  $Z = 2$, lattice results give slightly lower abundances than that of the ideal gas.
This is suggestive of dimer-dimer ($nn$--$pp$ or $np$--$np$) interactions reducing the dimer abundance.
The observed discrepancies become significant as the values of $A$ and $Z$ increase.

For the case of the trimer, noticeable differences first appear for $A = 4$, which suggests that the trimer abundance is increased due to nucleon-trimer interaction.
We also observe a similar behavior in the case of the tetramer or alpha, and we find that the nucleon-alpha interactions similarly enhance the alpha abundance.
From the consistency of the results in Fig.~\ref{fig3} and Fig.~\ref{fig4}, we infer that the effects of cluster-nucleon and cluster-cluster interactions are responsible for the observed deviations from the ideal gas abundances.

In summary, we have used NLEFT simulations to investigate nuclear clustering in hot dilute nuclear matter. We have introduced a theoretical framework called light-cluster distillation to determine the abundances of light nuclear clusters.
The corresponding abundances of clusters are well determined for the cases where heavier cluster abundances are negligible.
The comparisons with ideal gas predictions show excellent agreement at low densities $\rho \simeq 0.001~{\rm fm}^{-3}$.
At higher densities, we observe deviations from ideal gas abundances due to cluster-nucleon and cluster-cluster interactions.  Since we perform fully non-perturbative \emph{ab-initio} calculations starting from bare nucleonic interactions,
our results contain many-body correlations to all orders and can thus serve as benchmarks for approximate calculations and models such virial expansions, statistical models, and transport models.  In future work, we will perform analogous studies using high-fidelity chiral nuclear forces combined with state-of-art many-body methods such as perturbative quantum Monte Carlo~\cite{Lu:2021tab} together with the rank-one operator method \cite{Ma2023ro} and wave function matching~\cite{Elhatisari:2022qfr}.


\begin{acknowledgments}
We are grateful for discussions with
Fabian Hildenbrand, Bing-Nan Lu, Yuan-Zhuo Ma, and Shihang Shen.
This work was supported in part by the European
Research Council (ERC) under the European Union's Horizon 2020 research
and innovation programme (grant agreement No. 101018170),
by DFG and NSFC through funds provided to the
Sino-German CRC 110 ``Symmetries and the Emergence of Structure in QCD" (NSFC
Grant No.~11621131001, DFG Grant No.~TRR110).
The work of UGM was supported in part by VolkswagenStiftung (Grant no. 93562)
and by the CAS President's International
Fellowship Initiative (PIFI) (Grant No.~2018DM0034).
The work of DL is supported in part by the U.S. Department of Energy (DE-SC0021152, DE-SC0013365, DE-SC0023658, SciDAC-5 NUCLEI Collaboration). The authors gratefully acknowledge the Gauss Centre for Supercomputing e.V. (www.gauss-centre.eu)
for funding this project by providing computing time on the GCS Supercomputer JUWELS
at J\"ulich Supercomputing Centre (JSC).
\end{acknowledgments}


\clearpage


\beginsupplement
\onecolumngrid

\section{Supplemental Materials}

\subsection{Lattice Hamiltonian with Wigner SU(4)-symmetric nuclear force}\label{sec:hamiltonian}
We define the Hamiltonian in a cubic box with lattice coordinates $\bm{n} = (n_x,n_y,n_z)$.
We employ an SU(4)-symmetric interaction, and our Hamiltonian reads
\begin{equation}\label{eq:hamiltonian}
  H = K + \frac{C_2}{2}\sum_{\bm{n}}:\tilde{\rho}^2(\bm{n}): + \frac{C_3}{6}\sum_{\bm{n}}:\tilde{\rho}^3(\bm{n}):,
\end{equation}
where $K$ is the kinetic term with nucleon mass $m = 938.9~{\rm MeV}$, and the colons indicate normal ordering.
The smeared density operator $\tilde{\rho}(\bm{n})$ is defined as
\begin{equation}\label{eq:sL}
   \tilde{\rho}(\bm{n}) = \sum_i\tilde{a}^+_i(\bm{n})\tilde{a}_i(\bm{n}) + s_L\sum_{|\bm{n}'-\bm{n}|=1}\sum_i\tilde{a}_i^+(\bm{n}')\tilde{a}_i(\bm{n}'),
\end{equation}
where $i$ is a collective spin-isospin index, and the smeared creation and annihilation operators are defined as
\begin{equation}
   \tilde{a}_i(\bm{n}) = a_i(\bm{n}) + s_{NL}\sum_{|\bm{n}'-\bm{n}|=1}a_i(\bm{n}').
\end{equation}
The summation over the spin and isospin implies that the interaction is SU(4) invariant.
Here, the nearest-neighboring smearing parameter $s_L$ controls the strength of the local part of the interaction,
while the nearest-neighboring smearing parameter $s_{NL}$ controls the strength of the nonlocal part of the interaction.
The couplings $C_2$ and $C_3$ give the overall strength of the two-body and three-body interactions, respectively.

In this work, we perform our calculations at a lattice spacing $a=1.32$~fm, which corresponds to a momentum cut-off $\Lambda=\pi/a\approx471~{\rm MeV}$, and we utilize averaged twisted boundary conditions~\cite{Lu2020thermal_SM}.
The parameters values of our interaction are set as $s_L = 0.061 $, $s_{NL} = 0.5$, $C_2 = -3.41\times10^{-7}~{\rm MeV}^{-2}$, and $C_3= -1.4\times10^{-14}~{\rm MeV}^{-5}$.
These parameters are proposed in Ref.~\cite{Lu2019Essential_SM} by fitting the properties of light and medium-mass nuclei. Moreover, it has been observed that this specific set of parameters effectively captures the thermal properties exhibited by nuclear matter~\cite{Lu2020thermal_SM}.

\subsection{Pinhole trace algorithm}
In canonical ensemble calculations, we fix the parameters of nucleon number $A$ and temperature $T$, yielding the expectation value of any observable $\mathcal{O}$ as follows,
\begin{equation}\label{eq:ensemble_average}
  \langle\mathcal{O}\rangle_\beta = \frac{Z_{\mathcal{O}}(\beta)}{Z(\beta)} = \frac{{\rm Tr}_A(e^{-\beta H}\mathcal{O})}{{\rm Tr}_A(e^{-\beta H})},
\end{equation}
where $Z(\beta)$ is the canonical partition function, $\beta=T^{-1}$ is the inverse temperature, $H$ is the Hamiltonian, and ${\rm Tr}_A$ is the trace over the $A$-nucleon states.
Throughout this work, we use units where $\hbar = c = k_B =1$.

In Ref.~\cite{Lu2020thermal_SM},  the pinhole trace algorithm (PTA) is proposed to evaluate the Eq.~\eqref{eq:ensemble_average}.
The explicit form of the canonical partition function $Z(\beta)$ is expressed in the single particle basis as follows,
\begin{equation}\label{eq:partition_function}
  Z(\beta) = \sum_{c_1,\cdots,c_A}\langle c_1,\cdots,c_A|\exp(-\beta H)| c_1,\cdots, c_A\rangle,
\end{equation}
where the basis states are Slater determinants composed of $A$ nucleons. $c_i=(\bm{n}_i,\sigma_i,\tau_i)$ denotes the quantum numbers of the $i$th nucleon, where $\sigma_i$ is the spin and $\tau_i$ is the isospin.
On the lattice, the components of $\bm{n}_i$ take integer numbers from 0 to $L-1$, where $L$ is the box length in lattice units.
The neutron number $N$ and proton number $Z$ are separately conserved, and the summation in
Eq.~\eqref{eq:partition_function} is limited to the subspace with specified values of $N$ and $Z$.
The inverse temperature $\beta$ is divided into $L_t$ slices with temporal lattice spacing $a_t$ such that $\beta = L_ta_t$, where $a_t = (2000~{\rm MeV})^{-1}$ is taken in this work.

We use the Hubbard-Stratonovich transformation~\cite{Str57_SM,Hubbard59_SM} and a discrete auxiliary field which couples to the nucleon density,
\begin{align}\label{eq:auxiliary_field}
  :\exp\left(-\frac{a_tC_2}{2}\tilde{\rho}^2-\frac{a_tC_3}{6}\tilde{\rho}^3\right):
  &  = \sum_{k=1}^D\omega_k :\exp\left[\sqrt{-a_tC_2}\phi_k\tilde{\rho}\right]:\nonumber\\
  & = \sum_{k=1}^D :\exp\left[\sqrt{-a_tC_2}(\phi_k\ln \omega_k)\tilde{\rho}\right]:,
\end{align}
where the $\omega_k$'s and $\phi_k$'s are real numbers, and we require $\omega_k> 0$ for all $k$.
Taking $D=3$ and expanding Eq.~\eqref{eq:auxiliary_field}  up to $\mathcal{O}(\tilde{\rho}^4)$,
the constants $\phi_k$'s and $\omega_k$'s can be determined by comparing both sides order by order, see Ref.~\cite{Lu2019Essential_SM} for details. Now we can rewrite  partition function $Z(\beta)$ in terms of the transfer matrix operator with discrete auxiliary field,
\begin{align}\label{eq:partition_function2}
  Z(\beta) = \sum_{c_1,\cdots,c_A}\int\mathcal{D}s_1\cdots\mathcal{D}s_{L_t}\langle c_1,\cdots,c_A|\,                  M(s_{L_t})\cdots M(s_1)|c_1,\cdots,c_A\rangle,
\end{align}
where
\begin{align}
  M(s_{n_t}) = :\exp\left[-a_tK + \sqrt{-a_t C_2}\sum_{\bm{n}}s_{n_t}(\bm{n})\tilde{\rho}(\bm{n})\right]:
\end{align}
is the normal-ordered transfer matrix, and $s_{n_t}$ is our shorthand for all auxiliary fields at $\bm{n}$, $n_t$~\cite{Lee2009PPNP_SM, Lahde2019book_SM}.
Note that for the decomposition in Eq.~\eqref{eq:auxiliary_field}, the $s_{n_t}$ can only take several discrete values $\{\phi_k\ln\omega_k\}$,
so the path integral over real variables means the summations over indices.
For a given configuration $s_{n_t}$, the transfer matrix $M(s_{n_t})$ consists of a string of one-body operators which are directly applied to
each single-particle wave function in the Slater determinant.
Similar to the previous work~\cite{Lu2020thermal_SM}, the abbreviations $\vec{c} = \{c_1,\cdots,c_A\}$ and $\vec{s} = \{s_1,\cdots,s_{L_t}\}$ are used for notational convenience.

To evaluate the Eq.~\eqref{eq:partition_function2}, the projection Monte Carlo methods with importance sampling are used to generate an ensemble
$\Omega$ of ${\vec{s}, \vec{c}}$ of configurations according to the relative
probability distribution
\begin{equation}
  P(\vec{s},\vec{c}) =|\langle\vec{c}|M(s_{L_t})\cdots M(s_1)|\vec{c}\rangle|.
\end{equation}
The expectation value of any operator $\hat{O}$ can be expressed as
\begin{align}
  \langle\hat{O}\rangle = \langle \mathcal{M}_O(\vec{s},\vec{c})\rangle_\Omega/\langle \mathcal{M}_1(\vec{s},\vec{c})\rangle_\Omega,
\end{align}
where
\begin{align}
  \mathcal{M}_O(\vec{s},\vec{c}) = &\langle\vec{c}|M(s_{L_t})\cdots M(s_{L_t/2+1}) \: \hat{O} \:
                                                        M(s_{L_t/2})\cdots M(s_1)|\vec{c}\rangle/P(\vec{s},\vec{c}).
\end{align}
To generate the ensemble $\Omega$,  $\vec{s}$ and $\vec{c}$ are updated alternately.
First for a fixed nucleon configuration $\vec{c}$ the auxiliary fields $\vec{s}$ are updated through the shuttle algorithm~\cite{Lu2019Essential_SM}, then the nucleon configuration $\vec{c}$ is updated using the Metropolis algorithm.
The details for updating $\vec{s}$ and $\vec{c}$ can be found in Ref.~\cite{Lu2020thermal_SM}.

\subsection{Correlation functions}

\begin{figure}[!htbp]
  \centering
  \includegraphics[width=0.65\textwidth]{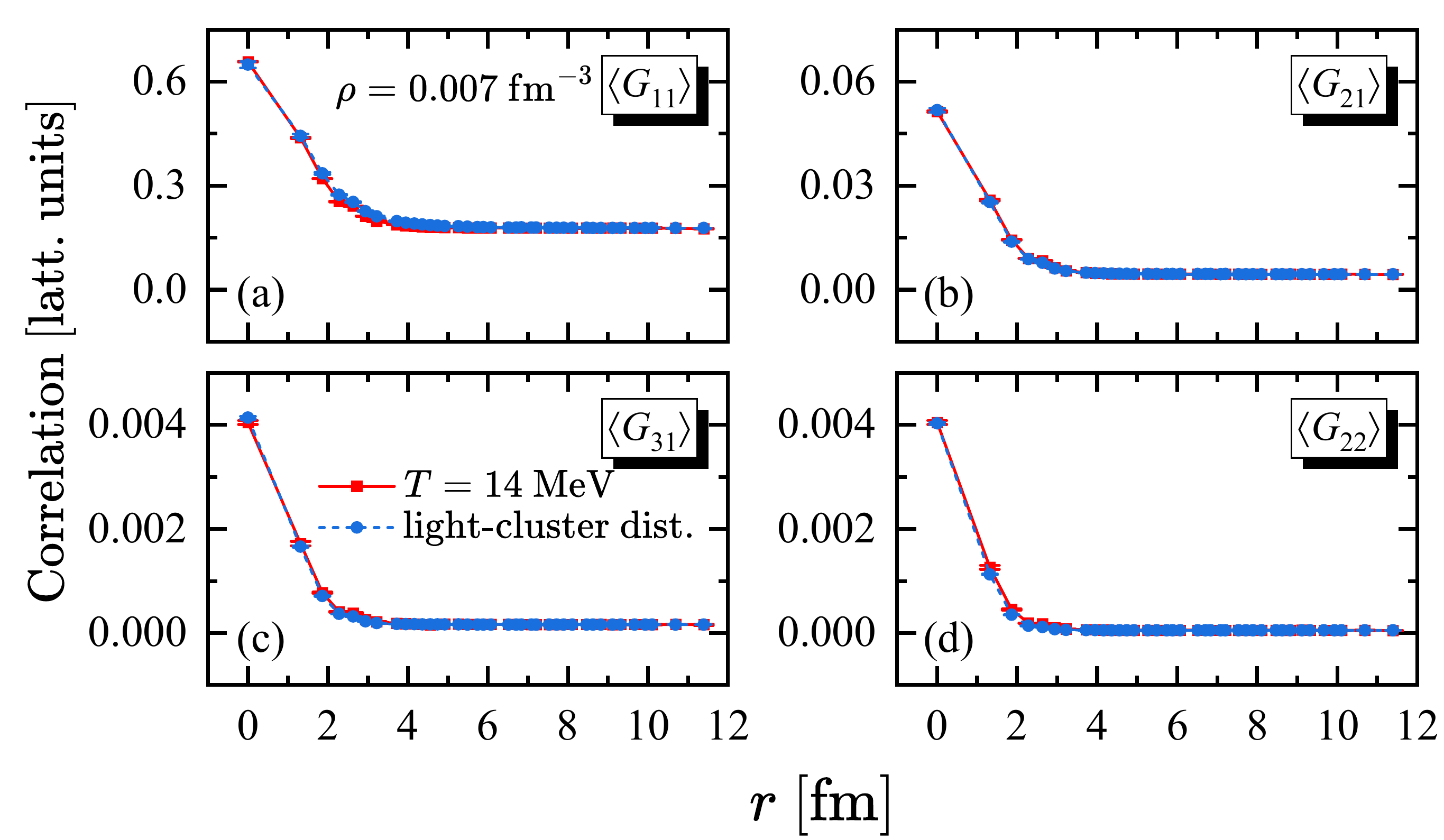}\\
  \caption{Correlation functions $G_{11}$, $G_{21}$, $G_{31}$, and $G_{22}$ for relative distance $r$ in units of $\rm fm$.
  The solid lines and red squares show lattice MC results at temperature of $T=14~{\rm MeV}$ using $A=16$ and $L = 10$ which corresponds to the density $\rho = 0.007~{\rm fm}^{-3}$. The dashed lines and blue circles show the fitted results of light-cluster distillation.
  The lines connecting the data points are intended to guide  the eye.
  }\label{figS1}
\end{figure}

\begin{figure}[!htbp]
  \centering
  \includegraphics[width=0.65\textwidth]{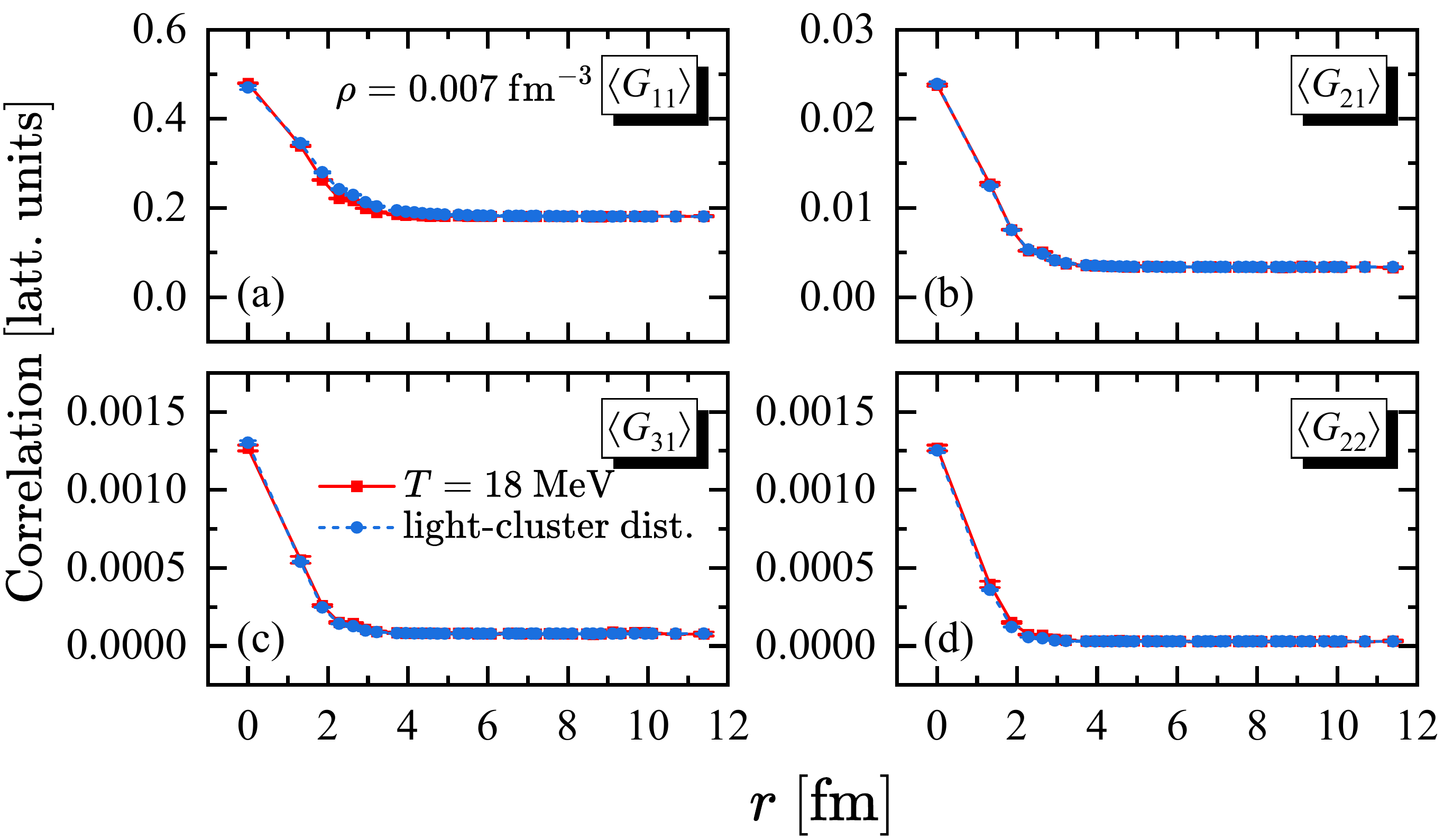}\\
  \caption{Correlation functions $G_{11}$, $G_{21}$, $G_{31}$, and $G_{22}$ for relative distance $r$ in units of $\rm fm$.
        The solid lines and red squares show lattice MC results at temperature of $T=18~{\rm MeV}$ using $A=16$ and $L = 10$ which corresponds to the density $\rho = 0.007~{\rm fm}^{-3}$. The dashed lines and blue circles show the fitted results of light-cluster distillation.
        The lines connecting the data points are intended to guide  the eye.
        .
  }\label{figS2}
\end{figure}

\begin{figure}[!htbp]
  \centering
  \includegraphics[width=0.65\textwidth]{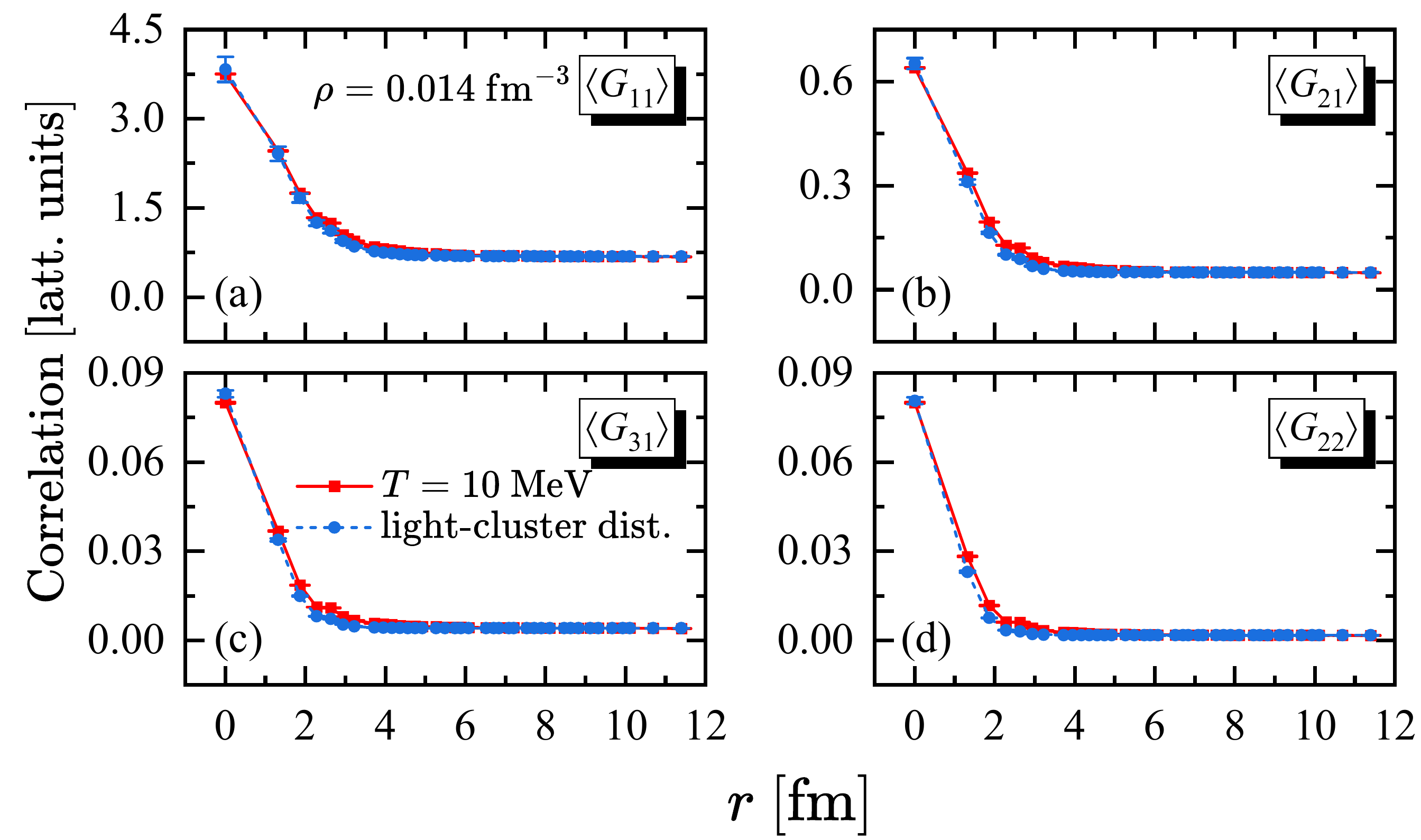}\\
  \caption{Correlation functions $G_{11}$, $G_{21}$, $G_{31}$, and $G_{22}$ for relative distance $r$ in units of $\rm fm$.
        The solid lines and red squares show lattice MC results at temperature of $T=10~{\rm MeV}$ using $A=32$ and $L = 10$ which corresponds to the density $\rho = 0.014~{\rm fm}^{-3}$. The dashed lines and blue circles show the fitted results of light-cluster distillation.
        The lines connecting the data points are intended to guide  the eye.
  }\label{figS3}
\end{figure}

\begin{figure}[!htbp]
  \centering
  \includegraphics[width=0.65\textwidth]{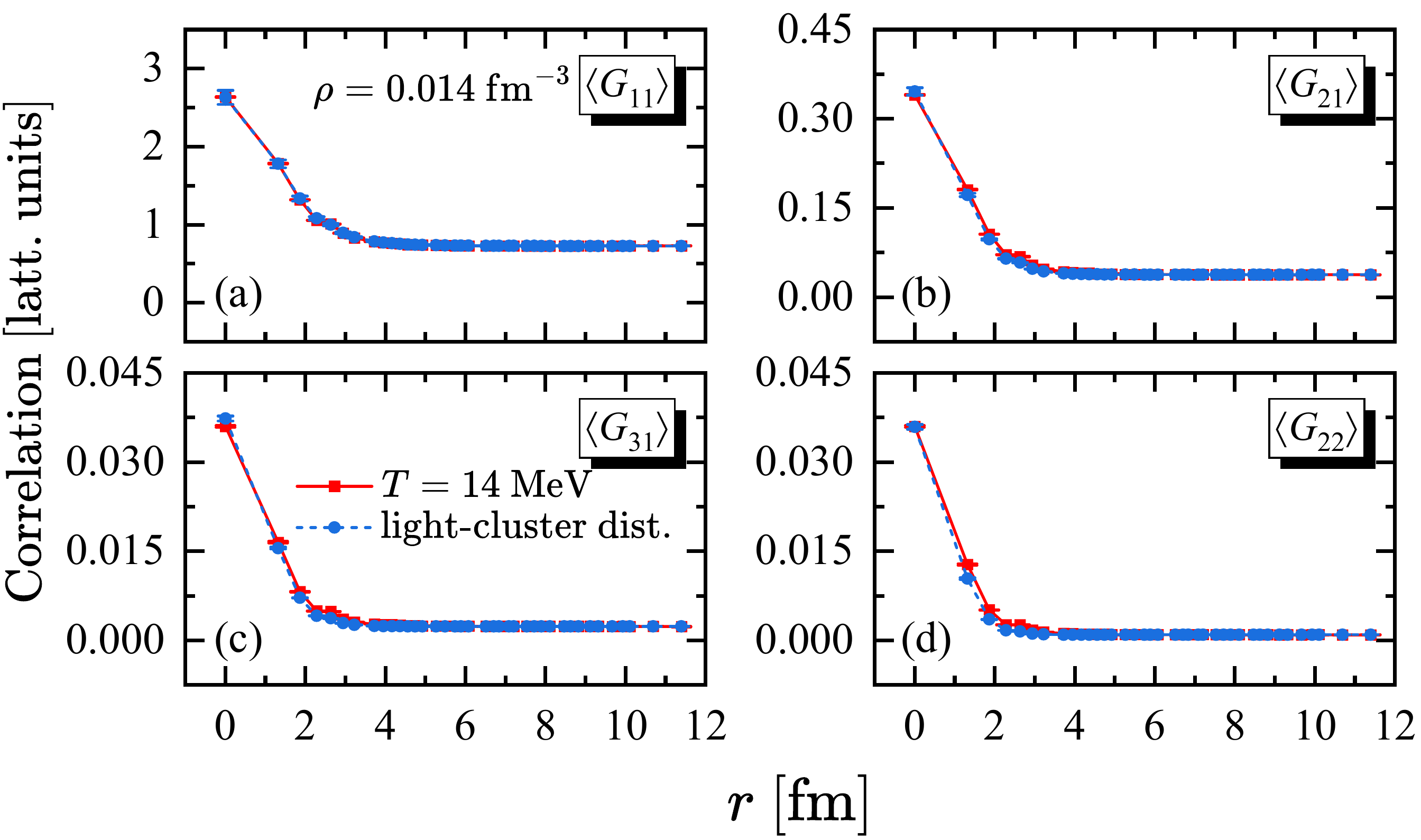}\\
  \caption{Correlation functions $G_{11}$, $G_{21}$, $G_{31}$, and $G_{22}$ for relative distance $r$ in units of $\rm fm$.
        The solid lines and red squares show lattice MC results at temperature of $T=14~{\rm MeV}$ using $A=32$ and $L = 10$ which corresponds to the density $\rho = 0.014~{\rm fm}^{-3}$. The dashed lines and blue circles show the fitted results of light-cluster distillation.
        The lines connecting the data points are intended to guide  the eye.
  }\label{figS4}
\end{figure}

\begin{figure}[!htbp]
  \centering
  \includegraphics[width=0.65\textwidth]{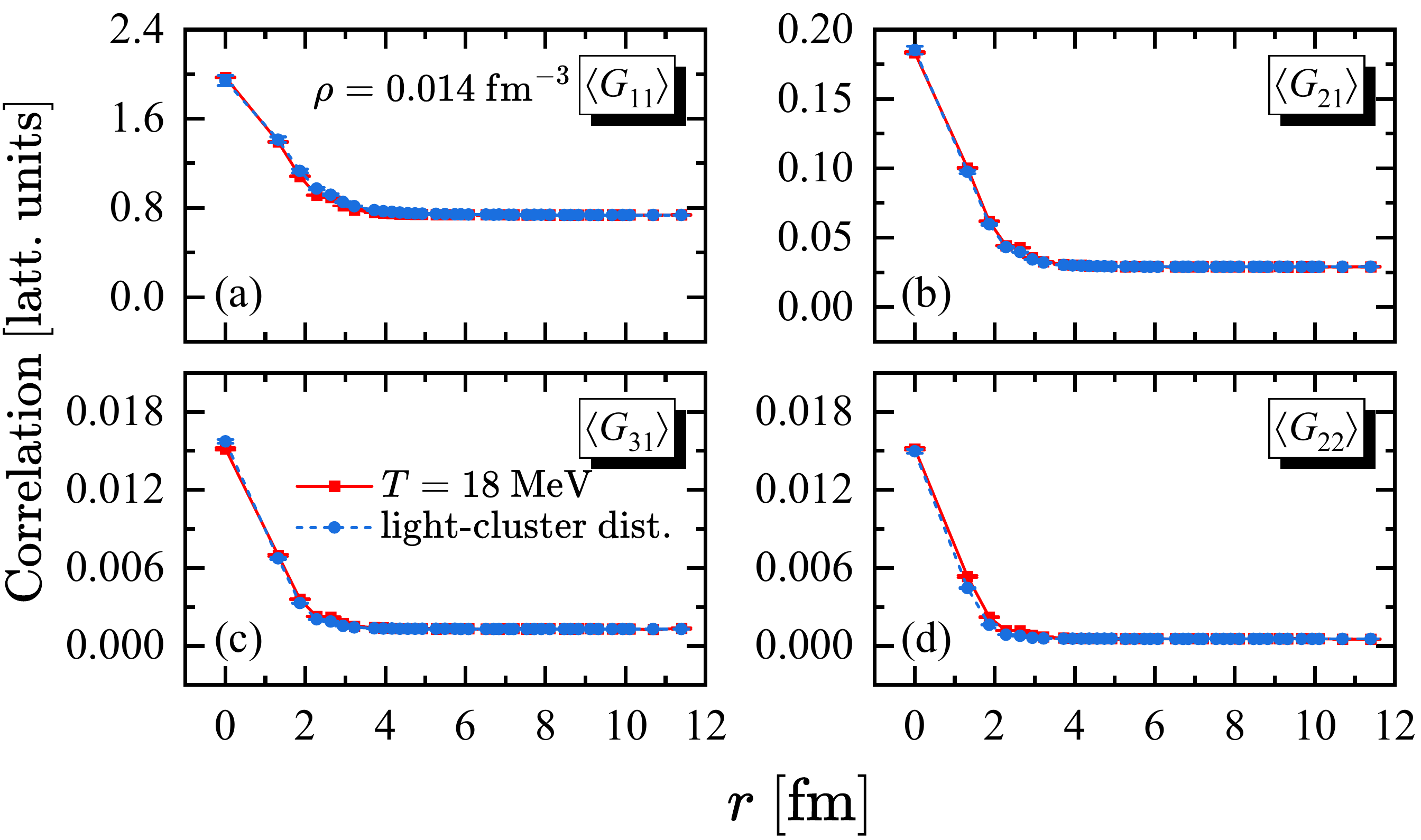}\\
  \caption{Correlation functions $G_{11}$, $G_{21}$, $G_{31}$, and $G_{22}$ for relative distance $r$ in units of $\rm fm$.
        The solid lines and red squares show lattice MC results at temperature of $T=18~{\rm MeV}$ using $A=32$ and $L = 10$ which corresponds to the density $\rho = 0.014~{\rm fm}^{-3}$. The dashed lines and blue circles show the fitted results of light-cluster distillation.
        The lines connecting the data points are intended to guide  the eye.
  }\label{figS5}
\end{figure}

In this section, we present the correlation functions $G_{11}$, $G_{21}$, $G_{31}$, and $G_{22}$ alongside the results from light-cluster distillation for various temperatures and densities.
Figs.~\ref{figS1} and \ref{figS2} depict the results for a density of $\rho=0.007~{\rm fm}^{-3}$ at temperatures of $T=14$ and $18~{\rm MeV}$, respectively. Also, Figs.~\ref{figS3}, \ref{figS4}, and \ref{figS5} show the results for a density of $\rho=0.014~{\rm fm}^{-3}$ at temperatures of $T=10$, $14$, and $18~{\rm MeV}$, respectively.
At $T=10~{\rm MeV}$ and $\rho=0.014{\rm fm}^{-3}$, we observe a noticeable decline in the agreement of light-cluster distillation for $G_{ij}$ due to the non-negligible fractions from heavier clusters with $A>4$.
As the temperature increases, the contribution of heavier clusters becomes less significant, resulting in an improved agreement between light-cluster distillation and the correlations.

\begin{figure}[!htbp]
  \centering
  \includegraphics[width=0.450\textwidth]{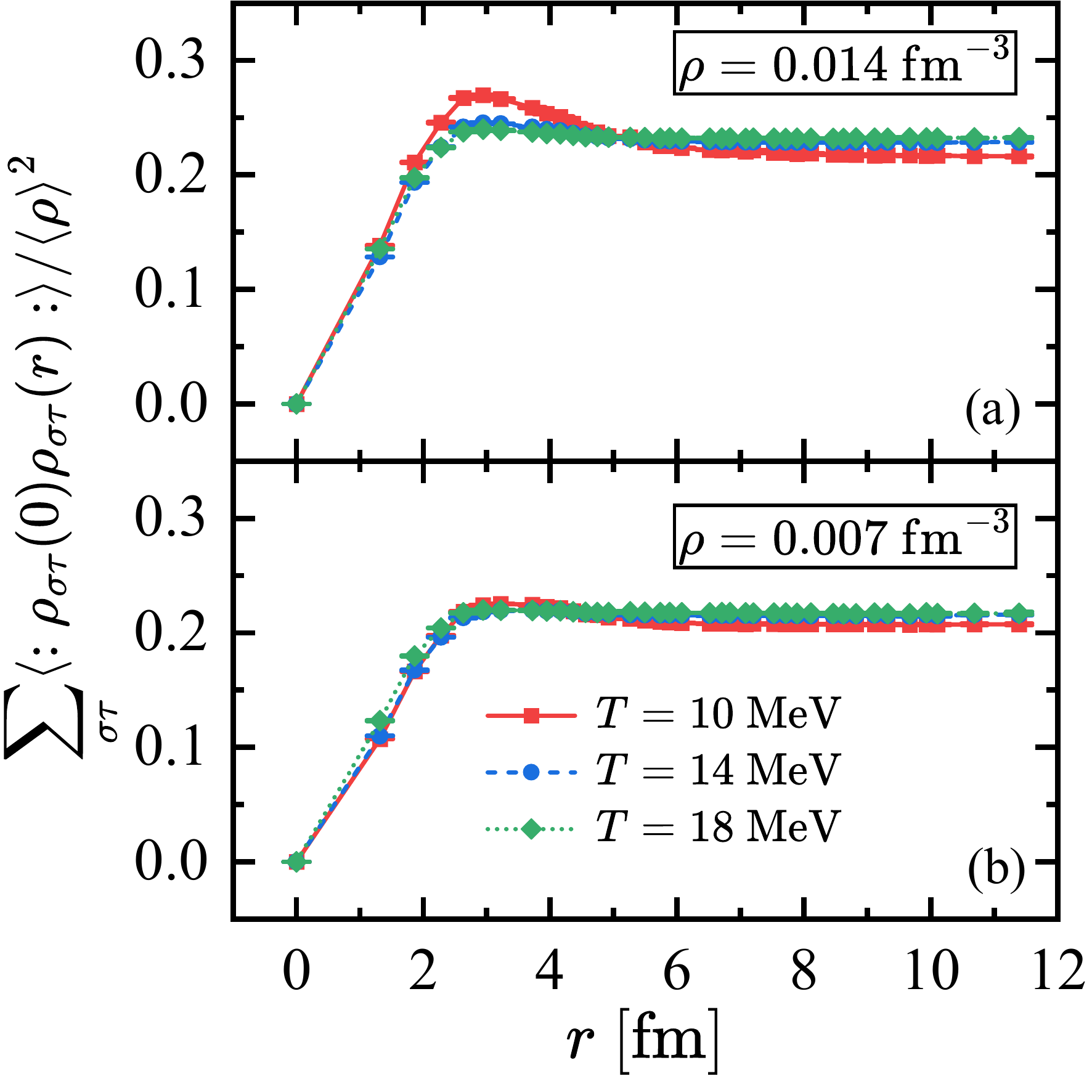}\\
  \caption{The correlation function $\sum_{\sigma\tau}\langle:\rho_{\sigma\tau}(0)\rho_{\sigma\tau}(\bm{r}):\rangle$ scaled by the square of density $\langle\rho\rangle^2$ as a function of relative distance $r$ in units of ${\rm fm}$.
   The top and bottom panels respectively show the results for the density $0.014$ and $0.007~{\rm fm}^{-3}$.
   The squares, circles, and diamonds correspond to the temperatures $10$, $14$, $18~{\rm MeV}$, respectively.
  }\label{figS6}
\end{figure}

Considering that heavier clusters induce spatial localizations of two nucleons with the same spin and isospin, they can be probed by examining the correlation function $\sum_{\sigma\tau}:\rho_{\sigma\tau}(0)\rho_{\sigma\tau}(\bm{r}):$.
Fig.~\ref{figS6} illustrates this correlation as a function of distance at three different temperatures.
The correlation function is zero at the origin due to the Pauli principle and increases up to its maximum value at around $r=3~{\rm fm}$.
For $\rho = 0.007~{\rm fm}^{-3}$, at all temperatures the correlation function remains nearly constant for larger $r$.
However, for $\rho=0.014~{\rm fm}^{-3}$, a significant peak is observed at temperature of $T=10~{\rm MeV}$, which provides a direct evidence for the emergence of heavier clusters.
The diminishing importance of heavier clusters with increasing temperatures  is further substantiated by the attenuation of this peak.

\subsection{Finite volume effects}
\begin{figure}[!htbp]
  \centering
  \includegraphics[width=0.45\textwidth]{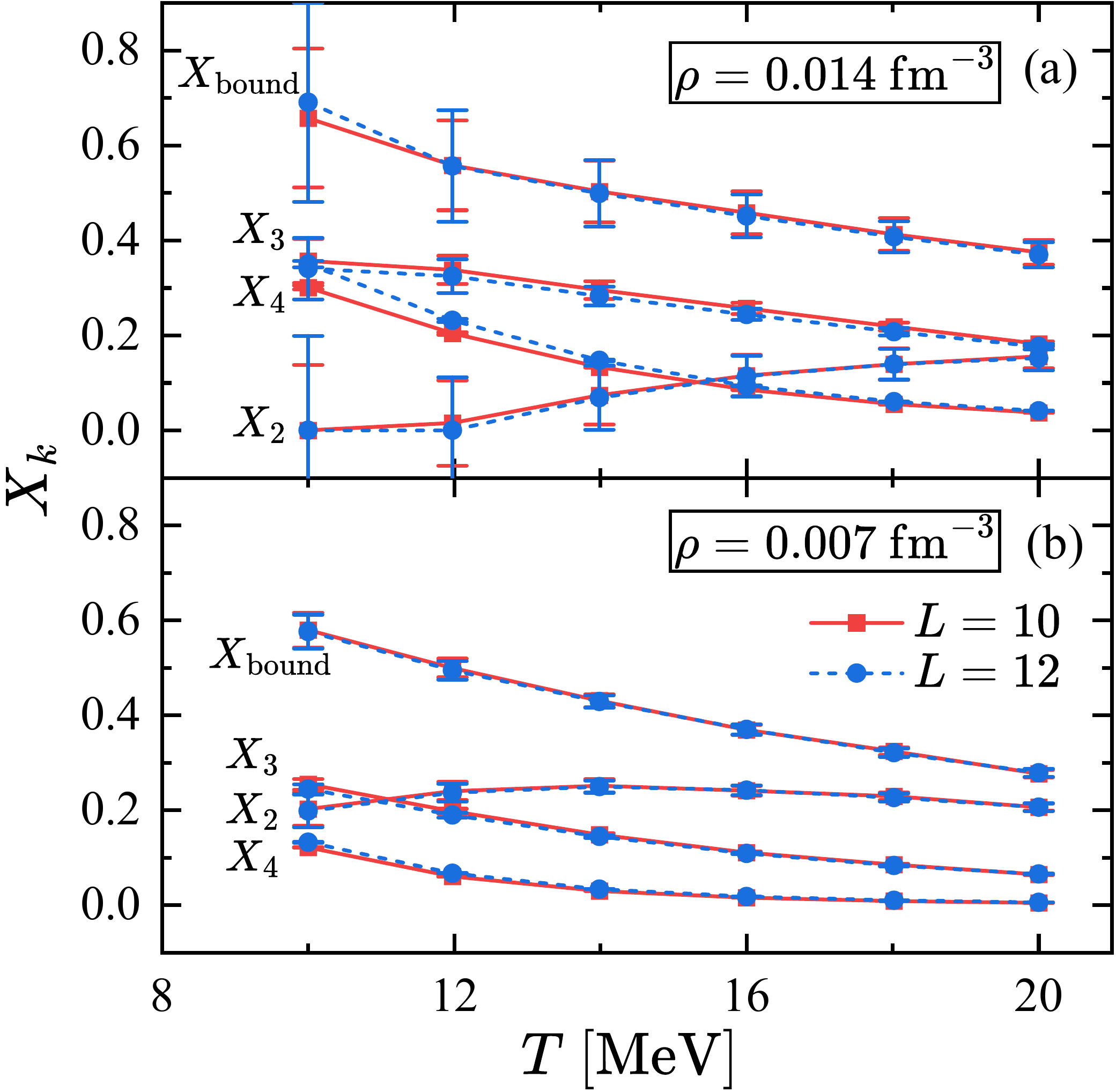}\\
  \caption{Comparison of the mass fraction $X_k$ from the box size $L=10$ (squares with solid lines) with the ones for $L=12$ (circles with dashed lines).
  Panels~(a) and (b) show the results of the densities $\rho = 0.014$ and 0.007~${\rm fm}^{-3}$, respectively.
  The error bars reflect the fitting uncertainty for the $w_k$ in light-cluster distillation.
  }\label{figS7}
\end{figure}
To check  the impact of finite volume effects on the mass factions $X_k$'s,
a larger box size of $L=12$ is used in the calculations. Total nucleon numbers are adjusted to $A=28$ and $56$ in order to achieve the densities of $0.007$ and $0.014~{\rm fm}^{-3}$, respectively.
In Fig.~\ref{figS7}, a comparison between the results obtained with $L=10$ and $L=12$ is shown as a function of temperature.
For the density $\rho = 0.007~{\rm fm}^{-3}$, almost perfect consistency is observed for all shown temperatures.
Some discrepancies of the center values are observed for $\rho = 0.014~{\rm fm}^{-3}$ at temperatures of $10$ and $12$~MeV due to the heavier clusters.
Nevertheless, good consistency is still found with increasing temperature, indicating that the obtained $X_k$ values from light-cluster distillation are quite reliable once the fractions of heavier clusters are negligible.

\subsection{Ideal gas estimated from the grand canonical ensemble}
\begin{figure}[!htbp]
  \centering
  \includegraphics[width=0.45\textwidth]{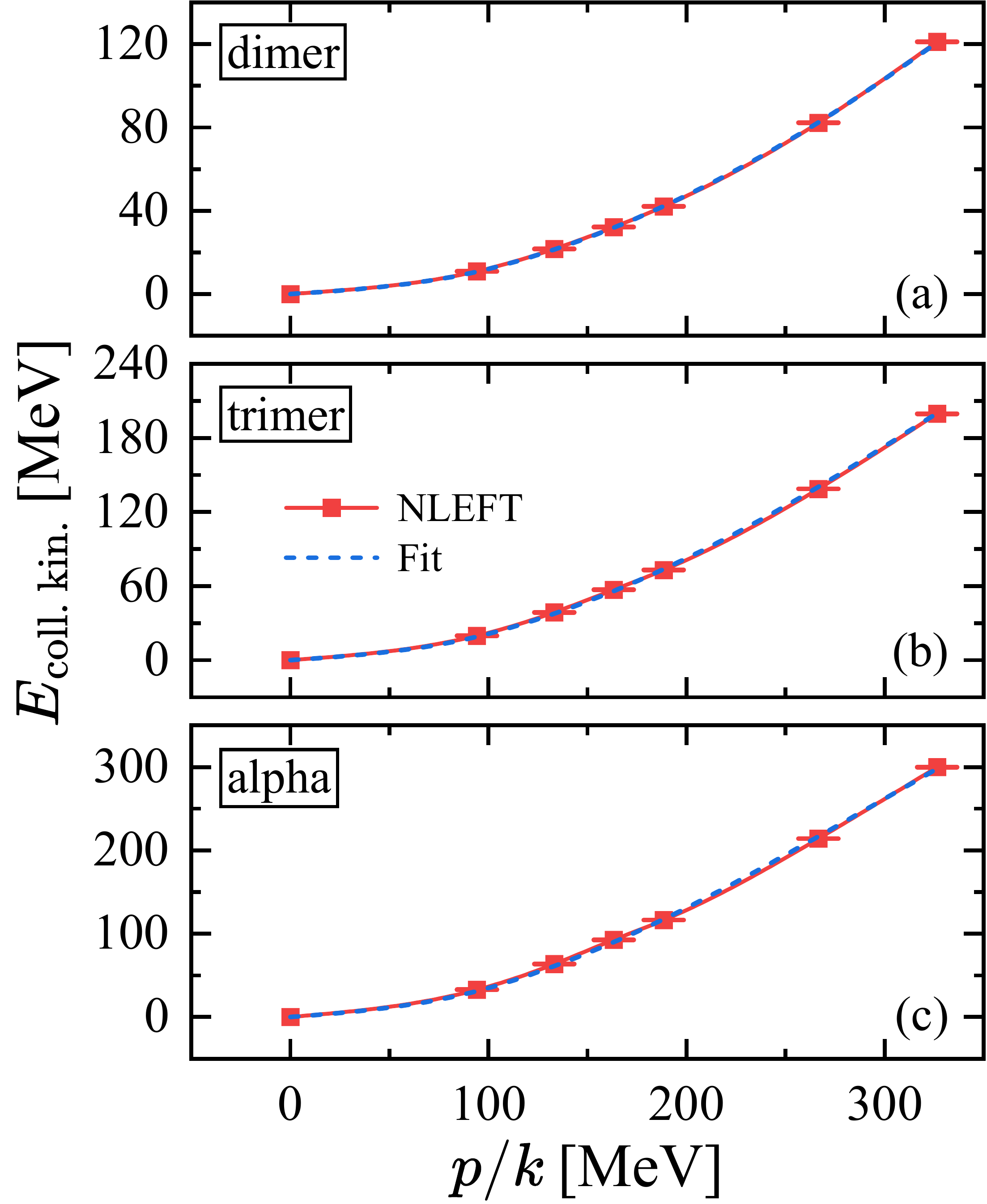}\\
  \caption{Dispersion relation of dimer (top), trimer (mid), and alpha (bottom) as a function of momentum scaled by the mass number.
  The squares represent the results from NLEFT, and dashed lines show the fitting results form Eq.~\eqref{eq:dispersion}.
  }\label{figS8}
\end{figure}

The system is approximated as an ideal gas composed by noninteracting nucleons, dimers, trimers, and tetramers (alphas).
We simply denote the corresponding chemical potentials as $\mu_1$, $\mu_2$, $\mu_3$, and $\mu_4$ with considering that the interactions in our lattice calculations are SU(4) invariant and the calculations are restricted to symmetric nuclear matter.
In the thermodynamic equilibrium with a temperature $T$, the average number for each species of particle are described via,
\begin{equation}\label{eq:N_A}
  N_k = \sum_{\varepsilon_k}g_k \frac{1}{e^{(\varepsilon_k-\mu_k)/T}+\eta}, ~~~\mbox{with}~k=1,2,3,4,
\end{equation}
where $g_k$ is the degeneracy from spin and isospin.
For nucleons, $g_1=4$.
For dimers, $g_2=6$, where $2$ from $nn$ and $pp$ dimers, and $4$ from $np$ dimers.
For trimers, $g_3=4$ from $nnp$ and $npp$ trimers.
For alpha, $g_4=1$.
The value $\eta=(-1)^{k+1}$ gives the Bose-Einstein ($k=2,4$) or Fermi-Dirac ($k=1,3$)  distributions.
The energy $\varepsilon_k$ is written as,
\begin{equation}\label{eq:energy}
  \varepsilon_k = E_k+ E_{\rm coll. kin.}(p),
\end{equation}
which contains the contributions from the ground state energy $E_k$ and collective kinetic energy $E_{\rm coll. kin.}(p)$ as a
function of the momentum $p$.
The $E_k$'s of the clusters are given by lattice calculations directly, while $E_{\rm coll. kin.}(p)$ of a cluster does not follow the exact dispersion
relation since the Hamiltonian Eq.~\eqref{eq:hamiltonian} contains a nonlocal interaction which breaks Galilean invariance.
In practice, we find that $E_{\rm coll. kin.}(p)$ is fitted with the  following formula,
\begin{equation}\label{eq:dispersion}
   E_{\rm coll. kin.}(p) = a_1\frac{p^2}{2km} + a_2\left(\frac{p^2}{2km}\right)^2.
\end{equation}
This can be seen from Fig.~\ref {figS8}, where $E_{\rm coll. kin.}(p)$'s on the lattice are obtained from boosted ground-state wave functions.

The summation over energy in Eq.~\eqref{eq:N_A} can be transferred into integration over momentum and spatial spaces,
\begin{equation}\label{eq:sum_int}
   \sum_{\varepsilon} = \frac{1}{(2\pi)^3}\int d^3p~d^3r=\frac{V}{2\pi^2}\int_0^\infty p^2dp.
\end{equation}
The average number $N_k$ could thus be expressed as,
\begin{equation}
  N_k = V\,\frac{g_k}{2\pi^2}\int_0^{\infty}dp~\frac{p^2}{e^{[\varepsilon_k(p)-\mu_k]/T}+\eta},
\end{equation}
The corresponding density $\rho_k$ reads,
\begin{equation}
    \rho_k = \frac{N_k}{V} = \frac{g_k}{2\pi^2}\int_0^{\infty}dp~\frac{p^2}{e^{[\varepsilon_k(p)-\mu_k]/T}+\eta}.
\end{equation}

For a given total nucleon density $\rho$ and temperature $T$, various density $\rho_k$'s satisfy the condition,
\begin{equation}
  \rho_1+2\rho_2+3\rho_3+4\rho_4 = \rho.
\end{equation}
In the thermodynamic equilibrium, the chemical potential of the cluster, $\mu_k$, is $k$ times of $\mu_1$, that is, $\mu_k = k\mu_1$.
Combined with the condition from $\rho$, we could determine $\mu_1$, and thus $\rho_k$ for each particles.
In this way, we could obtain the fractions $X_k=k\rho_k/\rho$.

\subsection{Ideal gas estimated from the canonical ensemble}
The system is still approximated as an ideal gas composed by noninteracting nucleons, dimers, trimers, and alphas,
while the total proton and neutron numbers are fixed to $Z$ and $N$.
Since now $Z$ and $N$ are not required to be same, we should explicitly distinguish $8$ species of particles, and denote their numbers as $N_k$ $(k=1,...,8)$ for proton ($k=1$), neutron ($k=2$), $pp$ dimer ($k=3$), $pn$ dimer ($k=4$), $nn$ dimer ($k=5$), $ppn$ trimer ($k=6$), $pnn$ trimer ($k=7$), and alpha ($k=8$).
They should separately fulfill proton and neutron number conditions,
\begin{subequations}\label{eq:NZ_number}
   \begin{align}
     &N_1 + 2N_3 + N_4+2N_6+N_7+2N_8 = Z,\\
     &N_2 + N_4 + 2N_5 + N_6 + 2N_7 + 2N_8 = N.
   \end{align}
\end{subequations}
For a given temperature $T$ and volume $V$, the relative probability for the configuration $\{N_k\}$ reads,
\begin{equation}
  P(\{N_k\}) = \prod_{k=1}^8 \left[ z_k^{N_k}Q_k^{}(N_k,V,T)\right],
\end{equation}
where $z_k=e^{\mu_k/T}$ is fugacity depending on chemical potential $\mu_k$ and $Q_k(N_k)$ is canonical partition function with $N_k$ free particles in a volume $V$~\cite{Pathria2011book_SM}.
Considering the fact that system is in equilibrium, the proton and neutron number conditions Eq.~\eqref{eq:NZ_number} could further simplify $P(\{N_k\})$ as
\begin{equation}
    P(\{N_k\}) = z_1^Zz_2^N\prod_{k=1}^8 \left[Q_k(N_k,V,T)\right].
\end{equation}
The average number $\langle N_k\rangle$ can be calculated by summing over all configurations fulfilling Eq.~\eqref{eq:NZ_number},
\begin{equation}
  \langle N_k\rangle = \frac{\sum_{\{N_k\}}N_kP(\{N_k\})}{\sum_{\{N_k\}}P(\{N_k\})}.
\end{equation}
Obviously, the chemical potential thus plays no role in determining $\langle N_k\rangle$.

The calculation of the canonical partition function is rather cumbersome, while the grand partition function $\mathcal{Q}$ is rather simple~\cite{Pathria2011book_SM},
\begin{equation}
  \mathcal{Q}(z,V,T) = \left\{
  \begin{aligned}
  &\prod_\varepsilon\left(\frac{1}{1-ze^{-\varepsilon/T}}\right)^g\qquad\mbox{Bose-Einstein distribution}~,\\
  &\prod_\varepsilon\left(1+ze^{-\varepsilon/T}\right)^g\qquad~~\mbox{Fermi-Dirac distribution}~,
  \end{aligned}
  \right.
\end{equation}
with $g$ the degeneracy factor.
With the integration in Eq.~\eqref{eq:sum_int}, $\mathcal{Q}(z,V,T)$ is evaluated as,
\begin{equation}
  \mathcal{Q} = \exp[\ln \mathcal{Q}] = \exp\left\{\eta g\frac{V}{2\pi^2}\int_0^{\infty}dp\cdot p^2\ln\left[1+\eta ze^{-\varepsilon(p)/T}\right]\right\},
\end{equation}
where $\eta=-1~(+1)$ for Bose-Einstein (Fermi-Dirac) distribution and energy $\varepsilon(p)$ is taken as in Eq.~\eqref{eq:energy}.
With the fugacity expansion of $\mathcal{Q}$,
\begin{equation}
    \mathcal{Q}(z,V,T) = \sum_{N=0}^{\infty}z^N Q(N,V,T),
\end{equation}
we note that canonical partition function $Q(N,V,T)$ can be obtained via the $N$th derivative of $\mathcal{Q}(z,V,T) $ over the
fugacity $z$ at the point $z=0$,
\begin{equation}
   Q(N,V,T) = \left.\frac{1}{N!}\frac{d^N}{dz^N}\mathcal{Q}(z,V,T)\right|_{z=0}.
\end{equation}
In practice, the above derivative is computed numerically.


\end{document}